\documentclass[twocolumn,showpacs,aps,prl,floatfix,superscriptaddress]{revtex4-1}
\usepackage{amsmath,amssymb,eucal}
\usepackage{graphicx}
\usepackage{float}
\usepackage{multirow}

\begin{document}

\title{Olejarz, Krapivsky, Redner, and Mallick Reply}

\author{Jason Olejarz}
\affiliation{Center for Polymer Studies, and Department of Physics, Boston University, Boston, MA 02215, USA}
\author{P.~L.~Krapivsky}
\affiliation{Department of Physics, Boston University, Boston, MA 02215, USA}
\author{S.~Redner}
\affiliation{Center for Polymer Studies, and Department of Physics, Boston University, Boston, MA 02215, USA}
\author{K.~Mallick}
\affiliation{Institut de Physique Th\'{e}orique CEA, IPhT, F-91191 Gif-sur-Yvette, France}

\begin{abstract}
\end{abstract}


\maketitle

In Ref.~\cite{Letter}, we investigated a two-dimensional interface that grows
in one octant of the cubic lattice.  Exploiting limiting cases and symmetry,
we conjectured two basic nonlinear equations of motion for the interface
speed.  Combining these equations allows us to fit the interface speed along
the $(1,1,1)$ diagonal perfectly, but this fitting requires an unnatural
choice of fitting parameters.  Aesthetics suggests that one of our elemental
equations,
\begin{equation}
\label{zt-3d-final}
z_t =\frac{z_x}{z_x-1}\, \frac{z_y}{z_y-1}
\left[1-\frac{1}{z_x+z_y}\right]\equiv R\,,
\end{equation}
accurately describes corner $3d$ interface growth.  While the prediction from
\eqref{zt-3d-final}, $w=0.125$, for the interface speed in the $(1,1,1)$
direction accurately matches our measured value $w=0.1261(2)$, small
discrepancies persist.  The comment~\cite{SR} studies the same interface
growth rules starting from a flat interface perpendicular to $(1,1,1)$.
This work finds $w=0.12606(2)$, consistent with our numerics, but also in
slight disagreement with the solution to the conjectured exact equation.



We recently found other independent equations that satisfy the required
symmetries.  One example is
\begin{equation}
\label{zt_n}
z_t =R\times \frac{(1-z_x-z_y)^n}{1+(-z_x)^n+(-z_y)^n}\,
\end{equation}
for arbitrary $n$.  Setting $n=1+\log_3(8w)$ perfectly matches numerics; for
$w=0.12606$, $n=1.0077$.  However, the conjecture \eqref{zt-3d-final} is
aesthetically more compelling.

It seems coincidental that the beautifully symmetric growth equation
\eqref{zt-3d-final} should differ from simulations by less than a percent.
It is also unsatisfying to reproduce the numerics with high accuracy by using
an equation such as (10) in our original Letter or \eqref{zt_n}, which
contain unnatural fitting parameters.  We speculate that systematic effects
in the simulations may generate small discrepancies with the prediction of
\eqref{zt-3d-final}.
Also note that little is known analytically about long-lived transients in
$(2+1)$-dimensional Kardar-Parisi-Zhang growth \cite{KPZ_rev}.  A recent
numerical study \cite{HH} reveals a similarly stubborn approach to
asymptotics in measurements of scaling exponents for KPZ growth models.  On a
similar note, it is conceivable that differences in height correlation
functions between the flat hypercube-stacking model \cite{HS_model} of the
comment and the curved corner interface that we examine may generate slight
differences in the interface speed $w$.  Such differences in interfacial
statistics between flat and curved geometries have been found rigorously in
analogous $(1+1)$-dimensional growth models \cite{PS}.

While we agree with \cite{SR} that the numerics slightly deviate from the
predictions of \eqref{zt-3d-final}, it seems rash to reject this simple
equation of motion in favor of an unnaturally complex one that minutely
improves the accuracy for the interface speed.  The outstanding challenge, of
course, is to derive the correct equation for the interface motion.

\end{document}